\documentstyle[12pt]{article}
\newcommand{\beq}{\begin{equation}}
\newcommand{\eeq}{\end{equation}}
\newcommand{\eqna}{\begin{eqnarray}}
\newcommand{\eqne}{\end{eqnarray}}
\newcommand{\dia}{\begin{displaymath}}
\newcommand{\die}{\end{displaymath}}
\newcommand{\eqnaa}{\begin{eqnarray*}}
\newcommand{\eqnae}{\end{eqnarray*}}
\def\H{Hamiltonian }

\def\rf{(\ref}
\def\Sch{Schr\"odinger }
\def\Gh{\hat{G}}
\def\qh{\hat{q}}
\let\ALTsect=\section
\def\section{\setcounter{equation}{0} \ALTsect}

\begin{document}
\sloppy
\begin{center}
{\Large \bf
Body Fixed Frame, Rigid Gauge Rotations \\
and  Large N Random Fields in QCD}\\
\vskip .5cm
{\large{\bf Shimon Levit}\\}
\vskip .5cm
{\it Department of Physics}\\
{\it Weizmann Institute of Science}\\
{\it Rehovot 76100 Israel\ \footnote{ Permanent address. Supported in
part by
US -- Israel Binational Science Foundation grant no. 89--00393}}\\
{\it and}\\
{\it Max-Planck-Institut f\"ur Kernphysik}\\
{\it D-6900 Heidelberg, Germany\ \footnote{ Supported by Humboldt Award.}}\\
\end{center}
\vskip .5cm
\centerline{\Large{\bf Abstract}}

 The "body fixed frame" with respect to  local gauge transformations is
 introduced.  Rigid gauge "rotations" in QCD and their \Sch equation
 are studied for static and dynamic quarks.
Possible choices of the rigid gauge field configuration
 corresponding to a nonvanishing
static colormagnetic field in the "body fixed" frame are discussed.
A gauge invariant variational equation is derived in this frame.
For large number  N of colors
 the rigid gauge field configuration is regarded as random
with maximally random probability distribution under constraints on
macroscopic--like quantities. For the uniform magnetic field
the joint probability distribution of the field components
is determined  by maximizing the
appropriate entropy under the area law constraint for the Wilson loop.
In the quark sector the gauge invariance requires
the  rigid gauge field configuration to appear
not only as a background  but  also  as inducing
an instantaneous quark-quark interaction.  Both are random in the large N
limit.
\vskip .5cm
\section{Introduction}

Studies of  non perturbative aspects of dynamics
of non abelian gauge fields will continue to remain one of
the focuses of  theoretical activities. These fields appear
at all levels of the "elementary" interactions and even begin to
enter at a more phenomenological macroscopic level in condensed matter
systems. Quantum Chromodynamics represents a prime example of a
strongly coupled theory with non abelian gauge fields.  Despite many
efforts, e.g. instantons \cite{ins}, large N expansion
\cite{lar,fra,wit,mgm,egk,ind},
lattice gauge theory and  strong coupling
expansion \cite{lat,str}, topological considerations \cite{pol,top},
 QCD sum rules \cite{svz}, "spaghetti" vacuum \cite{cop},  light cone
approach \cite{bro}, explicit color projection \cite{jon}, and others
\cite{adl,yaf,dua}, the quantitative understanding
of the basic QCD features is still far from satisfactory. A sustained
effort with different angles of attack is clearly in order with the hope
that accumulated qualitative experience will finally lead to the
development of quantitative calculational tools. This paper is a
contribution to this effort.

 Invariance under local gauge transformations is the most important
feature of a non abelian gauge theory. In the framework of the Hamiltonian
formulation of QCD I wish to explore the consequences of this
invariance using some
general methods common to molecular and nuclear physics. I wish to
define an appropriate generalization of the body-fixed
 (intrinsic, rotating) frame formalism in the context of
the local gauge transformations. After doing so one can attempt to
separately investigate the
dynamics of the gauge "rotations" of the frame and the intrinsic frame
dynamics. These would be the analogs of rigid rotations and
intrinsic vibrations in molecular and nuclear physics. Most of my
interest in this paper will concentrate on the study of the "rigid
gauge rotations".
The study of the couplings between the "rotations" and
the "vibrations" of the gauge field is deferred to future work.
To avoid misunderstanding I wish to stress that by "gauge rotations" or
"rotations of the gauge field"
in this paper I will always mean Eq.(\ref{rgd}) below, which  includes
the proper SU(N) rotation  as well as the inhomogeneous "shift" term.

Perhaps the most important conceptual advantage of using the body-fixed
frame associated with a given symmetry
lies in the fact that one can freely approximate the dynamics in this
frame without fears to violate the symmetry.
In particular the use of this  formalism appears to be
fruitful provided there exist such a body-fixed,  intrinsic
frame in which the "rotational - vibrational" coupling can be considered
as small. This in turn generically
happens when the "rotational inertia" is much
larger than the "inertia" associated with the intrinsic motion
so that a variant of the Born-Oppenheimer approximation is valid.  A typical
situation is when the system's ground state is strongly "deformed" away
from a symmetric state.
By deformation I mean absence of symmetry with respect to
"intrinsic transformations",
i.e. transformations in the body-fixed frame  and not
with respect to the transformations in "laboratory". Absence of symmetry
in "laboratory" would correspond to the symmetry breakdown
which can not occur for a local gauge symmetry.
Quantum mechanical examples of deformed bodies are e.g. non spherical
molecules, deformed nuclei, etc.

I do not have a priori arguments that the QCD vacuum
is strongly "deformed" in the above sense. Appearance of various QCD
condensates, Ref. \cite{svz}, suggests  that this may be true. The
condensate wavefunction should then play a role of the strongly deformed
configuration. Another positive indication is the large N master field concept,
Ref. \cite{wit}, according to which a special gauge field  configuration
should exist which dominates the vacuum wave function or the
corresponding functional integral. It is expected,  however, that the
master  field is not
simply a fixed classical configuration. It should rather be regarded as
a statistical distribution allowing to calculate quantities which are
analogous to macroscopic
thermodynamic quantities in statistical physics, i.e. such that their
fluctuations are suppressed
in the large N limit, Refs. \cite{mat,ran}. Glimpses of the meaning of
these vague notions were found in various matrix models,
cf. Refs. \cite{mgm,egk,mat,ran},
and, e.g. in 1+1 dimensional QCD, Ref. \cite{bor}.  If this view point
is correct then suitably chosen rigidly rotating "deformed" gauge field
could play a role of the master field provided one understands
in which sense it should also be statistical.
The following formalism will clarify some of these issues and
provide a general framework in which they could be further discussed.

Works in the spirit of our study have already appeared in the
past,cf. Refs. \cite{lee,jac,sim,dos} and the analogy with various
types of rotational motion  is frequently used in QCD.
In this sense the present study is a continuation of these works.

This paper is organized as following. In Section 2
I introduce the transformation to the body-fixed frame in the context
of  the simplest model of the gauge rotational motion  --
the rigid gauge rotor.             Giving a  natural definition
of the rigid gauge "rotations" I  proceed to determine
the appropriate generalization of the standard space rigid rotor
 results --  the moment of inertia tensor,
the "body-fixed"
frame, the generators of the "body-fixed" gauge group
in terms of which the character of "deformation" can be classified, etc.
Despite severe limitation on the set of the allowed gauge field
configurations the model is gauge invariant.
 I work out in Section 3
the quantum mechanics of the  model. As with the
space rotor the generators of the "laboratory" and "body-fixed"
gauge transformations provide a complete set of quantum numbers
for the wave functionals of the model. The vacuum has zero energy and
is the most disordered state. Higher states correspond to the presence
of very heavy, i.e. static quarks and antiquarks in the system.
As an important example I consider the wave function and the
corresponding \Sch  equation  for  a pair of static quark and antiquark.
This and other similar equations in the model are
simple matrix equation with the inverse "moment of inertia" determining
the interaction between the color sources and depending on the assumed
rigid gauge configuration which plays the role of the "free parameter".

In Section 4 I discuss the meaning of the results obtained so far and
possible choices of the rigid gauge field which physically represents
a non vanishing  colormagnetic field
in the body-fixed  frame. In the ground state this frame
does not "rotate" but has random orientations in local
color spaces at every space point. Introduction of static quarks forces
the frame to "rotate" quantum mechanically
at the points where the quarks are situated.  The energy eigenvalues
of these "rotations" are the energies of the quantum states of the
colorelectric field generated by the  quarks. The propagator of
 this field
is the moment of inertia of the model and depends explicitly on the
assumed configuration of the rigid static colormagnetic field.
For zero field the propagator is a simple Coulomb while for
a uniform field diagonal in color the
propagator  behaves asymptotically as a decaying Gaussian.
  The so called dual Meissner effect picture of the confinement,
 Ref. \cite{man}, could be implemented  if a configuration
of the rigid colormagnetic field is found which "channels" the
colorelectric field and makes its propagator effectively one dimensional.
It turns out that the creation of such a magnetic "wave-guide"
is connected with existance of a zero eigenvalue of a certain
operator in the model.

Since  the quark color degrees of freedom are treated
 quantum mechanically
the model allows for a possibility that confinement of fundamental
representations  does not automatically mean confinement of higher
representations.   I discuss this
possibility and derive a variational equation for the rigid field.
This equation is fully gauge invariant.

In Section 5
expecting that  rigid gauge rotations should be relevant for the
master field concept I study the model in the large N limit.
Any candidate for the master field must be allowed to undergo free
"gauge rotations" which can not be frozen by this limit
and should induce an interaction between the quarks.
Going to the body fixed frame of these rotations I regard the
rigid gauge field configuration as random and introduce a natural
requirement that it is least biased under  constraints that it
should reproduce
gauge invariant quantities which can be regarded macroscopic-like in the
large N limit. This means that it should be maximally random under these
constraints. In order to make these ideas explicit I discuss in some
detail the case of the uniform
colormagnetic field. Such configuration in QCD was already discussed in the
past, Refs. \cite{sav,cop} but it seems that its appearance in the interaction
is a novel feature
of the model. The detailed form of  this interaction depends on the
differences of the color components of the magnetic field. It is not
confining for any finite number of colors. For $N \rightarrow \infty$
 I assume  that the form of the density of the color components of
  the field is known.
In 2+1 dimensions I choose  it such that it gives area
law for space oriented Wilson loops.
I treat then the entire distribution of these components as
 a joint distribution of their probabilities and regard
 the adopted "single component" density as an analog of a
macroscopic quantity that must be reproduced by this
joint distribution. I postulate that it must otherwise be maximally random,
 i.e.  must have the maximum entropy (minimum
information content) under suitable constraints. In this way I
derive the maximally random distribution for this model. I discuss its relation
to the large N limit of the \Sch  equation for the static quark--antiquark
system. I also give possible generalizations of this development
to  3+1 dimensions.

In Section 6 I include dynamical quarks
and show that the rigid gauge rotor limit corresponds exactly to the
limit in QED in which only the instantaneous Coulomb interaction
between the charges is retained. The major difference in QCD is that
in this limit   the quarks not only interact instantaneously via a more
complicated interaction  controlled by the rigid gauge field configuration,
but at the same time are also found  in a static  colormagnetic field
induced by this configuration. This dual appearance of the rigid field is
 a consequence of the gauge invariance and in the large N limit
is apparently the way the
master field should enter the quark sector of the theory. According to the
ideology developed in Section 5 both the field in which the quarks
move and their interaction should be considered as random in the large N
limit. The random interaction between the quarks  opens interesting
possibilities to discuss the relationship between confinement and
localization.

The body fixed  Hamiltonian with dynamical quarks is gauge invariant.
Its invariance  with  respect to global symmetries however is not
guaranteed for an arbitrary choice of the rigid gauge configuration.
I discuss possible variational approaches to determine this configuration
and derive an analogue of the Hartree-Fock equations for the model.

In the rest of the Introduction I will establish my notations, cf.,
 Ref.\cite{bjo}.  I
consider the QCD Hamiltonian in d=3 space dimensions in the $A^{0} = 0$
gauge,
\beq
H=\frac{1}{2}\int d^{3}x [(E_{a}^{i}(x))^{2}+(B_{a}^{i}(x))^{2}]
+ \int d^{3}x q_{\gamma}^{+}(x)[\alpha^{i} \left(p^{i}-g
A_{a}^{i}(x) \frac{ \lambda_{\gamma \delta}^{a}}{2} \right)
+\beta m]q_{\delta}(x).
\label{ham}\eeq
with
\beq
B_{a}^{i}(x)=\epsilon_{ijk}(\partial_{j}A_{a}^{k}+gf_{abc}A_{b}^{j}A_{c}^{k}),
\eeq
$i,j,k=1,...,3$;  $\gamma,\delta$=1,...,N;  $a=1,...,N^2 - 1$ for SU(N)
gauge group and  $f_{abc}$ -- the structure constants of the SU(N).
Dirac and flavor indices are omitted and the summation
convention for all repeated indices is employed here and in the following.
The gluon vector potential  $A_{a}^{i}(x)$ and  minus the electric
field  $-E_{a}^i (x)$ are canonically conjugate variables,
\beq
[E_{a}^{i}(x),A_{b}^{j}(y)]=i\delta_{ab}\delta_{ij}\delta (x-y),
\eeq
and the quark fields obey the standard anticommutation relations.

The Hamiltonian (\ref{ham}) is invariant under
the time independent gauge transformations. Using the matrix valued
hermitian fields
\beq
A_{\alpha\beta}^{i}(x)=A_{a}^{i}(x) \frac{ \lambda_{\alpha\beta}^{a}}{2},
\; \; \; \; \;
E_{\alpha\beta}^{i}(x)=E_{a}^{i}(x) \frac{\lambda_{\alpha\beta}^{a}}{2},
\eeq
where $\lambda^{a}$ are the SU(N) generators with the properties
\eqna
[\lambda^{a},\lambda^{b}]&=&2if_{abc}\lambda^{c};\; \; \;
\{\lambda^a,\lambda^b\} = \frac{4}{N}
\delta_{ab} + 2d_{abc} \lambda^c;\nonumber \\
Tr\lambda^{a}\lambda^{b}&=&2\delta_{ab};\; \; \; \; \;
\lambda_{\alpha \beta}^{a} \lambda_{\gamma\delta}^{a} = 2[\delta_{\alpha\delta}
\delta_{\beta\gamma}-\frac{1}{N}\delta_{\alpha\beta}\delta_{\gamma\delta}]
\eqne
one can write the gauge transformation as
\beq
 A^i \rightarrow SA^{i}S^{+}+\frac{i}{g}S\partial^{i}S^{+};\;\; \;
E^i\rightarrow SE^{i}S^{+};\; \; \;  q\rightarrow Sq  \eeq
where S(x) are time independent but x - dependent unitary $N\times N$
matrices, elements of the SU(N) group. The generators of this
 transformation
\beq
 G_{a}(y) \equiv G_a^A (y) + G_a^q (y),  \nonumber \eeq
\beq
G_{a}^A (y) =  \partial_{i}E_{a}^{i}(y)+gf_{abc}A_{b}^{i}(y)E_{c}^{i}(y),
\; \; G_{a}^q (y) = -gq^{+}(y)
\frac{\lambda^{a}}{2} q(y) \label{gen} \eeq
are conserved,
\beq
\frac{\partial G_{a}(x)}{\partial t}=i[H,G_{a}(x)]=0\eeq
and it is consistent to impose the Gauss law constraints
\beq
G_{a}(x)|\Psi>=0 \label{gl} \eeq
for all physical states.
Although $G_{a}(x)$ do not commute, their commutators
\beq
[G_{a}(x),G_{b}(y)]=gf_{abc}\delta (x-y)G_{c}(x)\eeq
allow to set  them all simultaneously zero.

\section{Rigid Gauge Rotor.}

In this section I will discuss the rigid gauge "rotations". Classically
I define them as gauge field configurations of the type
\beq
 A^i(x,t) = U(x,t)a^{i}(x)U^{+}(x,t)+\frac{i}{g}U(x,t)\partial^{i}U^{+}(x,t)
\label{rgd}\eeq
where $a^i(x)$ are t-independent, fixed as far as their x-dependence is
concerned, "rigid" fields which I do not
specify and leave them arbitrary for the moment. Eq.(\ref{rgd}) is
the simplest example of the transformation to the "body fixed" frame
of the local gauge symmetry in which I have assumed that the dynamics
of the field in this frame is  very stiff so that     the field
can  be approximately replaced by
its static  average. In general $a^i$ is of course dynamical but should
be viewed as constrained since $U(x,t)$ already contains a third of the
degrees of freedom. For non rigid  $a^i$  there is no obvious choice of
the body-fixed frame and it can be constrained in a variety of ways,
say, $a^3 = 0$ (axial gauge), $\partial_i a^i = 0$ (Coulomb gauge),
etc. In our language
these different gauge fixings correspond to different "rotating" frames.
Since they are "non inertial" the dynamics will look very differently
depending on the choice of the frame and  different fictitious
forces, the analogue of  Coriolis and centrifugal forces,
will be present.  I am planning to discuss these issues elsewhere.

With the anzatz (\ref{rgd}) the covariant derivatives are
\beq
D^i \equiv \partial ^i - igA^i = U(x,t) d^i(x) U^{+}(x,t),\eeq
with fixed, rigid
\beq
d^i(x) \equiv \partial ^i - iga^i(x). \eeq
Inserting (\ref{rgd}) in the Hamiltonian (\ref{ham})
one finds that the gauge invariant potential term
$\sum_{i,a}(B_{a}^{i}(x))^{2} \sim \sum_{i,j}Tr[D^i,D^j]^2 = \sum_{i,j}
Tr[d^i,d^j]^2$ is
independent of the $U$'s, i.e. it is fixed, nondynamical in this model.
The dynamics of the gauge field
is governed by the kinetic energy, i.e. the term with
the electric field  in \rf{ham}).  Using
$\partial_0(U\partial_iU^{+}) = (i/2)U(\partial_i\omega)U^{+}$ with
$\omega = 2i U^{+}\partial_0 U$ one finds
\eqna
-E^{i} = \partial_0 A^{i} =  \frac{1}{2g}U [\omega,d^i] U^{+},
\label{dai}\eqne
and therefore the kinetic energy in (\ref{ham}) is
\beq
\frac{1}{4}\int d^3 x Tr(\partial_0 A^i)^2 =
 - \frac{1}{16g^2}\int d^3 x Tr\left( \omega[d^i,[d^i,\omega]]\right)
 \label{ke} \eeq
where I have disregarded surface terms, ignoring for the moment
possible non vanishing fields at infinity, non trivial topologies and
other global issues (cf.  below).

The double commutator in (\ref{ke}) with the summation over all indices,
$x,i$ and the color is the straightforward generalization of the
familiar double vector product summed over all particles indices
 in the moment of inertia tensor
appearing in the kinetic energy of rigid space rotations of system of
particles with fixed relative positions.
 Following this analogy  the energy (\ref{ke}) of the rigid gauge
 rotations can be written
\beq
E_{rot} = \frac{1}{4}\int d^3 x Tr(\omega I \omega) \label{en} \eeq
where the moment of inertia is defined as a differential matrix operator
such that
\eqna
I\omega &\equiv& - \frac{1}{4g^2} \left[ d^i,[d^i,\omega]\right] = \nonumber \\
&=& - \frac{1}{4g^2}\left(\partial_i^2 - ig[\partial_i a^i,\omega] -
2ig[a^i,\partial_i\omega] - g^2 \left[ a^i,[a^i,\omega]\right]\right) .
 \label{mom}
\eqne
 To obtain the corresponding
Hamiltonian one can use the gauge field part $G^A$ of the generators
(\ref{gen}). Using  Eqs. (\ref{rgd}), (\ref{dai}) and the
definition (\ref{mom}) one finds
\beq
G^A = [\partial^i - ig A^i,E^i] =
\frac{1}{2g}U \left[d^i,[d^i,\omega]\right] U^{+} = -2g U (I\omega) U^{+}.
\label{omg} \eeq
Defining the gauge generators in the rotating frame
$\Gh = U^{+} G^A U$, expressing $\omega = - I^{-1} \Gh /2g$ from \rf{omg}) and
substituting in (\ref{en}) one finds the Hamiltonian of the rigid gauge
rotor
\beq
H_{rot}^A = \frac{1}{16g^2}\int d^3 x d^3 y  Tr \Gh (x) I^{-1}(x,y)
\Gh (y) = \frac{1}{4g^2}\int d^3 x d^3 y \Gh_a(x) I_{ab}^{-1}(x,y)
\Gh_b(y).
\label{hrot} \eeq
where $I_{ab}^{-1}(x,y) = (1/4) Tr(\lambda^a I^{-1}(x,y)
\lambda^b)$ is proportional to the inverse of the operator $-d^i_{ac}
d^i_{cb}$ with $d^i_{ab} = \partial_i \delta_{ab} - g f_{abc} a^i_{c}$
and I assumed  that this operator  does not have zero eigenvalues.
In a more careful way of handling fields at infinity one should avoid
the integration by parts in (\ref{ke}). The  inverse "moment
of inertia" operator is then replaced by a less transparent
\beq
K_{aa'}(x,x') = \int d^3 y \left[ d^i_{bc}(y) I^{-1}_{ca}(y,x)\right]
\left[ d^i_{bc'}(y) I^{-1}_{c'a'}(y,x')\right].\eeq
Most of the following results remain valid for both forms of this
operator.

The meaning of the preceeding expressions is quite obvious. They are the
field-theoretic generalization of the standard rigid rotor results.
The unbroken
local gauge symmetry of QCD means that there are free SU(N) color
gauge "rotations" at
every space point. Expression (\ref{hrot}) shows that the "rotations"
at different  points as well as around different color axes are coupled
via the non diagonal
elements of the moment of inertia "tensor" $I_{ab}(x,y)$  in the manner
similar to the coupling between rigid rotations around different space axes in
systems of particles.

It does not seem to be useful to diagonalize the operator $I^{-1}$ in
(\ref{hrot}).
The standard diagonal form of the rigid rotor \H, i.e., $H = \frac{1}{2}\sum_a
L_a^2/I_a$  can be usefully achieved only in the case of
rotations corresponding to a single SU(2) group to which the familiar
rigid space rotations  belong. Diagonalizing the moment of inertia
in the case of higher groups will introduce combinations
of the generators multiplied by  matrices of orthogonal rotations.
These in general will not have
the group commutation relations.  Already for a single
SU(3) the group O(8) of
rotations in the adjoint space needed in order to diagonalize the
moment of inertia is much larger than SU(3).

 The actual values of the moment of inertia depend on the rigid
configuration $a^i (x)$ of the gauge field via the
expression (\ref{mom}). This comprises the "free parameter" of the rigid
gauge rotor model. For abelian theory or alternatively
in the limit $g \rightarrow 0$ the inverse of $I(x,y)$
appearing in Eq.(\ref{hrot})
is just the Coulomb propagator.
 In the opposite large $g$ or long wavelength limit
$I(x,y)$ becomes a local tensor  given by the last term in (\ref{mom})
which is obviously the SU(N)
generalization of the moment of inertia expression.

 An important feature of
the \H (\ref{hrot}) is that despite the severe limitation of the
allowed gauge field configurations imposed by  (\ref{rgd}) it remaines
gauge invariant. This is because (\ref{hrot}) depends on $\Gh$
rather than $G^A$. Under a gauge transformation $U \rightarrow SU$,
$G^A$ transforms as $SG^A S^{+}$ so that  $\Gh(x)$ and therefore
$H_{rot}^A$ stay invariant. The gauge invariance of \rf{hrot}) is the
simplest illustration of the usefulness of the introduction of the
body fixed frame. One can freely approximate the dynamics in this frame
without fears of violating the symmetry with respect to which the frame
 has been defined, i.e. the local gauge symmetry in the present case.

Consider another transformation, $U \rightarrow US$. Referring to
Eq.\rf{rgd}) one can interpret this transformation
either as the change of $U$ i.e. the transformation of the intrinsic
frame with respect to the rigid "shape" $a^i$ or
as the change of $a^i$ ,
 $a^i \rightarrow Sa^{i}S^{+}+\frac{i}{g}S\partial^{i}S^{+}$, i.e.
the  transformation of the intrinsic "shape" with respect
to the intrinsic frame.
Such transformations obviously form a group of local SU(N) gauge
transformations  which I will call
 the intrinsic or "body fixed"
  gauge group to distinguish it from the "laboratory" gauge
group of the ordinary gauge transformations. According to two different
interpretations of the intrinsic gauge transformations
given above one has two options. One option is to
 regard  $\Gh$'s as the generators of the intrinsic group.
They act on  the dynamical variables $U$ but they have a disadvantage in
that  the "laboratory"
group is not completely independent of such an  intrinsic group, e.g.
they both have identical Casimir operators. Another option is to formally
introduce operators which gauge transform  the intrinsic variables $a^i$.
They will have the same form as $G^A$'s but with $a^i$ replacing $A^i$.
Defined in this way the intrinsic group will be completely independent
of the "laboratory" gauge group but will act on nondynamical variables
which do not appear in the wavefunctions. Convenience should dictate
which one to use.

The above introduction of the intrinsic vs "laboratory" gauge groups
is obviously quite general with e.g. the definition of $\hat G(x)$ being
independent of the rigid rotor restrictions set by fixing $a^i$
to be nondynamical in \rf{rgd}).
Unlike the local gauge symmetry in "laboratory", the symmetry in the
"body fixed" frame can be broken.
E.g., the gauge invariant \H (\ref{hrot})
 is in general not invariant under the transformations of the intrinsic
gauge group. This is a simple example of the situation to which I
referred earlier as a possible existence of "deformation"
vs impossibility of the symmetry breakdown in the context of non
abelian local
gauge theory. The character of the deformation can be classified
using the intrinsic gauge group, e.g. in classification of
possible "deformed shapes" of the rigid gauge rotor \rf{hrot})
by the  transformation properties of the moment of inertia
$I_{ab}(x,y)$ under this group. Here I obviously adopt the second
interpretation of the intrinsic group. The invariance of
$I_{ab}(x,y)$ under all
intrinsic transformations would be analogous
to the spherical rotor limit  in the space rotation case.
The invariance under a continuous subgroup of the intrinsic group is the
analog of the axial symmetric rotor, etc.  Discrete intrinsic
subgroups should also be considered.

Consider a rigid gauge configuration $a^{i'}$ which is a gauge
transform of $a^i$, $a^{i'} = S(a^i +(i/g)\partial_i)S^{+}$.
  The Hamiltonian
\rf{hrot}) will have the same form with the same moment of inertia
but with $\Gh$ replaced by $S^{+}\Gh S$. The eigenvalues of this
transformed Hamiltonian will not change and will therefore depend only
on gauge invariant combinations of the rigid $a^i$, i.e. on the Wilson
loop variables $Tr P exp (ig\oint a^i dx_i)$.
\section{Static Quarks}

So far I have discussed the rigid gauge rotor limit
of only the first term in \rf{ham}). The resulting $H_{rot}^A$ is relevant
for the discussion of very heavy quarks.
They can be considered static as far as their translational motion is
concerned. They still have a wavefunction describing the motion of their color
degrees of freedom. Because of \rf{gl}) this motion is coupled to
the "rotations" of the gauge field which I will treat
using \rf{hrot}).

In the limit of $m\rightarrow \infty$  the quark kinetic energy
term $q^{+}\vec{\alpha}\vec{p}q$  and the quark color current coupling
$q^{+}\vec{\alpha}\lambda^{a}q$ in Eq.\rf{ham}) can be neglected
and the resulting Hamiltonian decouples into a part containing the gauge field
and another  containing the quarks, $H=H_{A}+H_{q}$, where
$H_{A}$ is the first term in (\ref{ham})
and $H_q = m\int d^3x q^{+} (x)\beta q(x)$.
The coupling  appears only via the Gauss law constraint, Eq.(\ref{gl}).
The wave function can not be taken
as a product $\Psi=\Psi(q)\Psi(A)$ but should be a local color singlet.
In the representation  in which
\[ \beta=\left( \begin{array}{cc}  1 & 0 \\  0 & -1\end{array}\right) \]
\[ q_{\alpha}(x)=a_{\alpha}(x) \left( \begin{array}{cc} 1 \\ 0\end{array}
\right)   + b_{\alpha}^{+}(x) \left( \begin{array}{cc} 0 \\ 1\end{array}
\right)  \]
with
\eqna
\{a_{\alpha}(x),a_{\beta}^{+}(y)\}&=&\delta_{\alpha\beta}\delta (x-y)
\nonumber \\
\{b_{\alpha}(x),b_{\beta}^{+}(y)\}&=&\delta_{\alpha\beta}
\delta (x-y), etc...\nonumber
\eqne
the eigenfunctions of $H_{q}$ are trivially written down.
Consider, e.g.,
\eqna
|\Psi(q)>\equiv |vac(q)> &=& |0>, \label{vac} \\
|\Psi(q)>\equiv |x_{0},\alpha> &=& a_{\alpha}^{+}(x_{0})|0> \label{1qu} \\
|\Psi(q)>\equiv |x_{0},\alpha ;y_{0},\beta > &=& a_{\alpha}^{+}(x_{0})b_{\beta}
^{+}(y_{0})|0> \label{2qu} \eqne
These wave functions describe respectively  zero quarks, one
static quark at $x_{0}$ with color component $\alpha$ and a static
quark - antiquark pair at $x_0$ and $y_0$. It is easy to form local
color singlets with these wave functions.
 For e.g. the quark-antiquark pair it is
\beq
 |\Psi>=\sum_{\alpha\beta} \Psi_{x_{0},\alpha ;y_{0},\beta}(A)|x_{0},
\alpha ;y_{0},\beta> \eeq
with the wave functional
$\Psi_{x_{0},\alpha ;y_{0},\beta}(A)$ satisfying
\begin{eqnarray}G_{a}^{A}(x)\Psi_{x_{0},\alpha ;y_{0},\beta }(A) =
g\delta(x-x_{0})\frac{\lambda_{\alpha\alpha^{'}}^{a}}{2}
\Psi_{x_{0},\alpha^{'} ;y_{0},\beta}(A)+g\delta(x-y_{0})
\frac{\bar{\lambda}_{\beta
\beta^{'}}^{a}}{2}\Psi_{x_{0},\alpha ;y_{0},\beta^{'}}(A)\label{cnd}
\end{eqnarray}
where $\bar{\lambda}_{\alpha\beta}^{a}=-\lambda_{\alpha\beta}^{a*}$.
The wave functional of the
gauge field should be a singlet at every point in space except
at the position of the quarks where it should transform as  N and $\bar{N}$
multiplets of SU(N).  This constraint together with the
\Sch  equation
\beq
H_{A}\Psi_{x_{0},\alpha;y_{0},\beta}(A)=E\Psi_{x_{0},\alpha;y_{0},\beta}(A)
\eeq
completely defines the problem for the gauge field.

 In the rigid gauge rotor limit $H_A$ is given by Eq.(\ref{hrot}).
The wavefunctions of this \H are general functionals
 $\Psi [U(x)]$ of the SU(N) matrices $U_{\alpha \beta} (x)$.
Their scalar product is determined by functional integration over the
$U$'s with the corresponding group invariant measure.
The vacuum wave functional must obey $G_a^A (x)\Psi_{vac}[U] = 0$.
This means that it is a constant independent of $U_{\alpha \beta}(x)$.
Since also $\Gh_a (x)\Psi_{vac}[U] = 0$ the vacuum energy is zero according
to (\ref{hrot}).
Regarding the parametrization of the $U$'s in terms of the appropriate Euler
angles of the SU(N) rotations at every space point, the constant
$\Psi_{vac}[U(x)]$ means that all the "orientations" of $U(x)$ at all points
are equally
probable, i.e. there are no correlations between the "orientations"
of the rigid gauge rotor at different points. This is as "random" as
the distribution of the $U$'s can get. The absence of correlations
is the property only of the vacuum. For other states the "orientations"
of the gauge fields at different space points are correlated via the
"moment of inertia" operator.

In order to discuss
the  wave functions with non zero number of quarks it is sufficient
to know some simple
properties of the gauge generators $G_a^A(x)$ and $\Gh_a(x)$. Since
$\Gh_a(x)$ are gauge scalars, they commute with $G_a^A (x)$,
\beq
[G_a^A(x),\Gh_b(y)] = 0 \eeq
which means that together the generators of the "laboratory" and the
intrinsic gauge groups provide a complete set of commuting
quantum numbers for the wave functionals $\Psi [U_{\alpha \beta }(x)]$.
Indeed, since the Casimir operators for $G^A$'s and $\Gh$'s coincide one has
e.g., for the SU(2) the $(G_a^A(x))^2$, $G_3^A(x)$ and $\Gh_3(x)$,
i.e. three local commuting operator fields  for the three fields of the
Euler angles needed to specify the $U(x)$.
In the SU(3) one has eight fields of the "Euler angles" and eight local
commuting generators made off $G^A$'s and $\Gh$'s --
 the two group Casimir operators, one Casimir operator
of an SU(2) subgroup for $G^A$'s, say $\sum_{a=1}^{3} (G_a^A)^2(x)$ and the
corresponding one for the $\Gh$'s and respectively two
pairs of the Cartan generators -- $G_3^A(x)$, $G_8^A(x)$ and $\Gh_3(x)$ and
 $\Gh_8(x)$. This counting continues correctly for any N, i.e.
$N-1$ for the SU(N) Casimir operators, $2((N-1)+(N-2)+... +1)$
for the  Casimir operators of pairs of SU$(N-1)$...SU(2) subgroups and $2(N-1)$
for the Cartan generators. Altogether there are $N^2 - 1$ local commuting
operators as needed. An eigenfunction
of this complete set of operators is  the Wigner function
$D_{K K^{\prime}}^L(U(x))$ of U at a certain space point. $K$ and $K^{\prime}$
are the quantum numbers of the "laboratory" and the intrinsic groups and $L$
determines the representation.

Since under an infinitesimal gauge transformations $U\rightarrow
(1+i\epsilon_a(x) \frac{\lambda_a}{2})U$  one can easily verify that
\eqna
\left[ G_{a} (x),U_{\alpha \beta}(y) \right] &=& g \delta (x - y)
\frac{\lambda^a_{\alpha \gamma}}{2} U_{\gamma \beta}(x) \nonumber \\
\left[ G_{a} (x),U_{\alpha \beta}^{+}(y) \right] &=& - g \delta (x -y)
U_{\alpha \gamma}^{+}(x)\frac{\lambda^a_{\gamma \beta}}{2} \label{com} \\
\left[ \Gh_{a} (x),U_{\alpha \beta}(y) \right] &=& g \delta (x - y)
U_{\alpha \gamma}(x) \frac{\lambda^a_{\gamma \beta}}{2} \nonumber \\
\left[ \Gh_{a} (x),U_{\alpha \beta}^{+}(y) \right] &=& - g \delta (x -y)
\frac{\lambda^a_{\alpha \gamma}}{2} U_{\gamma \beta}^{+}(x)
 \nonumber \eqne
 All the operators in the rigid gauge rotor model are functions of the
$G$'s and $U$'s. E.g. consider the electric field operator. According
to Eq.(\ref{dai}) it is
\beq
E^i =  -\frac{1}{4g^2} U [d^i,I^{-1} \Gh] U^{+} \label{elf} \eeq
where I have expressed $\omega$ in terms of $\Gh$ using (\ref{omg}).

Using the relations (\ref{com}) it is easy to write the general form of
the wave functions for a single quark and for a quark--antiquark pair,
\eqna
\Psi_{x_0,\alpha}[U] = U_{\alpha \gamma}(x_0)c_{\gamma} \nonumber \\
\Psi_{x_0,\alpha;y_0,\beta}[U] = U_{\alpha \gamma}(x_0)U_{\delta\beta}
^{+}(y_0)c_{\gamma \delta} \label{wfn}\eqne
They satisfy the conditions \rf{cnd}) following from the
Gauss law with constant coefficients
 $c_{\gamma}$ and $c_{\gamma \delta}$
which give  the probability amplitudes of the intrinsic quantum
numbers $\gamma$ and $\delta$. They should be normalized,
$\sum |c_{\gamma}|^2 = 1 ; \sum |c_{\gamma \delta}|^2 = 1$ to assure
the normalization $\int d[U(x)] |\Psi [U]|^2 = 1$

These amplitudes
 must be found by solving the corresponding \Sch equations but before
describing this I wish to remark that the above form of
the wave functions is valid also when the limitation of
the rigid gauge rotations is relaxed and the most general gauge
configurations are allowed. The parametrization \rf{rgd}) is still very
useful but now with fully dynamical fields $a^i$
 the variation of which should be limited only by a "gauge fixing"
condition to avoid overcounting as described above.
  The dynamics will of course be
that of the full QCD but  the wave functions
of the static quark and the quark--antiquark pair will have the same form
\rf{wfn}).  The difference will be that amplitudes $c_{\gamma}$ and
$c_{\gamma \delta}$ will be functionals of $a^i(x)$  describing the
space and
color fluctuations of the "string" attached to the quark or between the
quark and the antiquark. In the rigid gauge rotation case there are only
color fluctuations described by constant amplitudes.

For quarks in higher representations the wave functions have the same
form with $U$ replaced by the appropriate Wigner D-function. E.g. in
the adjoint representation
\eqna
\Psi_{x_0,a}[U] = Tr(U(x_0)\lambda^aU^{+}(x_0)\lambda^b)c_b,
\nonumber \eqne
etc.

I will now derive the \Sch  equation for the string amplitudes
$c_{\gamma}$ and $c_{\gamma\delta}$.
 Acting with the \H (\ref{hrot}) on \rf{wfn}), using (\ref{com})
and the orthogonality of $U$'s  with respect to the integration over the
group, $\int dU U_{\alpha \beta}^{*} U_{\mu \nu} = \delta_{\alpha \mu}
\delta_{\beta \nu}$,  I find
\eqna
Q_{\alpha \gamma}(x_0)c_{\gamma} =  E c_{\alpha}, \nonumber \\
Q_{\alpha \gamma}(x_0)c_{\gamma \beta} + Q^{*}_{\beta \mu}(y_0)c_{\alpha \mu}
- P_{\alpha \beta , \gamma \mu }(x_0,y_0)c_{\gamma \mu} =
E c_{\alpha \beta}, \label{sch1} \eqne
where I denoted
\eqna
Q_{\alpha \gamma}(x_0) = \frac{1}{4} I_{ab}^{-1}(x_0,x_0)(\lambda^a
\lambda^b)_{\alpha \gamma}, \\
P_{\alpha \beta , \gamma \mu }(x_0,y_0) =  \frac{1}{2} I_{ab}^{-1}(x_0,y_0)
\lambda^a_{\alpha \gamma} \lambda^b_{\mu \beta}.\eqne
In SU(2) $Q_{\alpha \gamma}$ takes a particularly simple diagonal form,
$Q_{\alpha \gamma} = \delta_{\alpha \gamma}(1/4)I_{aa}^{-1}(x_0,x_0)$.
and is the eigenvalue for a single quark. For quarks in e.g. adjoint
representation the lambda matrices in the expressions above are replaced
by the corresponding group generators $if_{abc}$.
The first two terms in the second line of
\rf{sch1}) are the quark and the antiquark self
energies whereas the last term is their interaction. In QCD one expects
that terms like $Q$ are inflicted by the long and short distance
divergences and should be properly regularized which I will assume for the
rest of the paper. I will further assume the translational invariance of
$Q$, i.e. its independence of $x_0$. One can then  rewrite Eq.\rf{sch1})
by transforming it to the basis in
which $Q$ is diagonal. Defining its eigenvectors
$Qb^{(n)} = \epsilon_n b^{(n)}$ and expanding
$c_{\gamma \beta} = d_{mn} b_{\gamma}^{(m)} b_{\beta}^{(n)*}$ one finds
\beq
<kl| P (x_0,y_0) |mn> d_{mn} = (E - \epsilon^k - \epsilon^l) d_{kl}
\;\;\; {\rm(no\;\;sum\;\;over\;\;k\;\;and\;\;l)} \label{peq} \eeq
where $<kl| P |mn> = P_{\alpha \beta , \gamma \mu } b_\gamma^{(m)}
b_\mu^{(n)*} b_\alpha^{(k)} b_\beta^{(l)*}$. The \Sch equation \rf{peq})
is $N^2 \times N^2$ matrix equation and the most interesting question of course
concerns the dependence of its eigenvalues on the distance $|x_0 - y_0|$
for various possible choices of the rigid gauge field configuration
$a^i(x)$ on which the matrix $P$ depends. I will address this question
in the next section.

\section{Choices of The Rigid Field. Mean Field Equations.}

The rigid configuration if it exists in QCD must reflect the
properties of the gluon condensate of the  vacuum. One of the more
accepted views of the QCD vacuum  is that this is a condensate of non
trivial topological configurations -- the Z(N) vortices,
c.f.,\cite{top}.  Although
such configurations are easily incorporated in the above formalism
I was not able to overcome
technical difficulties in working out a theory of their condensation.

On a heuristic level each Z(N) vortex carries
 a unit of flux of the colormagnetic
field. Condensation of the vortices presumably
means that there is a non zero average of this field in  the vacuum.
 Of course due to unbroken local gauge symmetry  it
must undergo free "gauge rotations" at each space point. In the ground
state this means that there are
equal probabilities of all the "orientations"   yielding  zero
average value in the laboratory. The finite   average
value of the condensate field
can only be "seen" in the "body fixed" frame and should appear in this
picture in the manner similar to
 $a^i$ in the expression (\ref{rgd}) for our rigid gauge
rotations. The field strength
\beq
B^i(x) = U(x)b^i(x)U^{+}(x),\,\,\, \  b^i = \frac{i}{g} \epsilon_{ijk}
 [d^j,d^k], \eeq
also averages to zero in the ground state but has a non zero value $b^i$
in the "body fixed" frame.

Via the dynamics of $U(x)$ the anzatz (\ref{rgd})  leads to colorelectric
field  (\ref{elf}) which propagates away from points
where $\Gh(x)$ is non zero, i.e. from  the location of static quarks.
The propagator of this field is controlled by the condensate
field $a^i$ which enters the expressions for $I^{-1}$ and $d^i$.
This propagator is a long range Coulomb potential for zero $a^i$ and
  is a Gaussian for
$a^i$ corresponding to a uniform colormagnetic $b^i$ (cf., below).
The screening of the propagation range of the colorelectric field
in the presence of the colormagnetic "condensate"  $b^i$ is
reminiscent of the dual to the Meissner effect of screening of
a magnetic field by the electric condensate of a superconductor.
This possibility of the dual Meissner effect is  of course a
standard scenario for confinement in QCD. It is expected that
tubes of flux of the colorelectric field are formed which
connect quarks and make their energy depend linearly on the distance.

In the present formalism a way to attempt to model the formation
of a confining string is to look for  such a configuration of
the rigid field $a^i(x)$ for which the propagator $I^{-1}$ behaves
roughly speaking  as
one dimensional for large separations along some  given  line in space
at the end of which quarks can be placed. This means that
a sort of magnetic "wave guide" should be constructed so that
the Green's function of the operator $-d^2_{ab} =
-d^i_{ac} d^i_{cb} =
 -(\partial_i \delta_{ac} - g f_{acc'} a^i_{c'})(\partial_i - g f_{cbb'}
 a^i_{b'})$ is
asymptotically  $\propto |x-x'|$ along, say, one of the coordinate
axes. In order to see the difficulties in finding such a configuration
consider for simplicity 2 space dimensions and choose
$a^1 = c(y)$ and $a^2 = 0$ with an arbitrary $c(y)$. This choice
corresponds to the colormagnetic field $b(y) = \partial_y c(y)$
depending only on one coordinate $y$. The operator to invert is then
\beq
-(\partial_x -ig c(y) \cdot F)^2- \partial_y^2   \eeq
where I denoted the color spin matrices $F^a_{bc} = if_{bac}$
and $c(y) \cdot F = c_a (y) F^a$. The propagator  is then
\beq
\int_{-\infty}^{\infty}
 dk e^{ik (x - x')} \sum_{n} \frac{\chi_n(k,y) \chi_n (k,y')}
{\epsilon_n (k)}, \label{prop} \eeq
where $\chi_n (k,x)$ and $\epsilon_n (k)$ are solutions of
\beq
 [-\partial_y^2 + (k +c(y) \cdot F)^2 ] \chi_n (k,y) = \epsilon_n (k)
\chi_n (k,y). \label{lan} \eeq
In order to achieve the desired confining behaviour of the propagator
the sum in (\ref{prop}) must be
$\sim k^2$ for $k \rightarrow 0$. The simplest
is to assume that the lowest eigenvalue of (\ref{lan}),
$\epsilon_0 (k)$ should vanish  as $k^2$ for small $k$. However
the operator in (\ref{lan}) is a sum of squares and does not have zero
eigenvalues for non trivial regular $c(y)$. It is also not symmetric
in $k$ for small values of $k$ but this seems to be less of  a problem.
The same conclusions seem to hold in 3 space dimensions. It is quite
possible that perhaps a singular configuration $a^i$ exists
which leads to  zero eigenvalue in (\ref{lan}) at zero $k$
but I was not able to find it.

The strong coupling limit of lattice QCD suggests
 that  quarks in the fundamental reperesentation are confined
whereas they are only
screened if put  in the adjoint representation. This
crucial difference  comes from fairly simple quantum mechanics
of color degrees of freedom related to  matching of group
representations in neighboring lattice points.
 In our rigid gauge rotor model a similar
simple quantum mechanics of  colors is retained. As a result the
eigenenergies of a systems of static quarks will be determined by
different
combinations of the color components of $I^{-1}$ depending on the
representation of the quarks.  E.g.,  as already mentioned
in Section 3  when the quarks are taken in
the adjoint representation the $\lambda$ matrices in the expressions
for $P$ and $Q$ in the \Sch equations (\ref{sch1})
are replaced by their adjoint counterpartners $F$.

In order to find the optimal $a^i$ in a systematic way  one can
follow a variational approach and  minimize the
ground state energy of the rigid gauge rotations for  fixed positions
of static quarks.
 This energy is given by a sum
of the lowest eigenenergy of $H^A_{rot}$, Eq.(\ref{hrot}) and the
colormagnetic energy given by the second term in (\ref{ham}) with
rigidly "rotating" $A(x)$, Eq.(\ref{rgd}), i.e.
\beq
E[a^i] = E_{rot}[a^i]  - \frac{1}{2g^2}\int d^3x Tr [d^i,d^j]^2 \eeq
Variation of this expression gives
\beq
\partial_i f^{ij} - ig [a^i,f^{ij}] = \frac{1}{2}\frac{\delta E_{rot}}
{\delta a^i} , \label{mfld} \eeq
where $f^{ij} = (i/g) [d^i,d^j]$ .
 Eq.(\ref{mfld}) is obviously gauge invariant.
 In the vacuum $E_{rot}$ is zero and the minimization of the second
term  simply gives the classical equation for $a^i$ in the vacuum.
For a quark- antiquark system $E_{rot}$ is non trivial and
depends on the distance between the quarks.  I plan to discuss
the solutions of the equation (\ref{mfld}) and their relation
to confinement elsewhere.

In the rest of this Section  as an illustration of a simple
 choice for the rigid field $a^i$ which allows
to obtain some analytic results I consider it  to be diagonal,
$a^i_{\alpha \beta} (x) = \delta_{\alpha \beta} a^i_\alpha (x)$.
The moment of inertia operator with such $a^i$ is
\beq
-\frac{1}{4g^2}\left[d^i ,[d^i ,\omega ] \right] _{\alpha \beta} =
-\frac{1}{4g^2} \left[ \partial ^i - ig(a^i_\alpha - a^i_\beta )\right] ^{2}
\omega_{\alpha \beta} . \label{dmi} \eeq
Using Green's function satisfying
\beq
\left(\partial ^i - ig(a^i_\alpha - a^i_\beta )\right) ^{2}
 J_{\alpha \beta }(x,y) = - \delta (x-y),\label{grf} \eeq
and following the procedure leading to Eq. (\ref{hrot}) one finds the rigid
gauge rotor Hamiltonian in this case
\beq
H = \frac{1}{4} \int d^2 x d^2 y \Gh_{\alpha \beta}(x) J_{\alpha \beta}(x,y)
\Gh_{\beta \alpha}(y). \eeq
The \Sch  equation for the static quark--antiquark wave function
\rf{wfn}) has the form \rf{peq}) with
\beq
Q_{\alpha \gamma}(x_0) = \frac{1}{4} \delta_{\alpha \gamma}\left[
\sum_\beta J_{\alpha \beta}(x_0,x_0) - \frac{1}{N}
\left(2 J_{\alpha \alpha}(x_0,x_0) -
\frac{1}{N}\sum_\beta J_{\beta \beta}(x_0,x_0)\right)\right] \eeq
\eqna
P_{\alpha \beta , \gamma \mu }(x_0,y_0) &=&- \frac{1}{2}
 \delta_{\gamma \mu} \delta_{\alpha \beta} J_{\alpha \gamma}(x_0,y_0) + \\
&+& \frac{1}{2N}\delta_{\alpha \gamma} \delta_{\mu \beta}\left[J_{\beta \beta}
(x_0,y_0) + J_{\gamma \gamma}(x_0,y_0)
 - \frac{1}{N} \sum_\nu J_{\nu \nu}(x_0,y_0)\right]. \nonumber \eqne
The diagonal components of $J$ are simple Coulomb propagators independent
of the color so that the expressions for $Q$ and $P$ can be simplified
 further but I will not go into the details of this. Instead
I will now consider the choice of $a^i$
which corresponds to a much discussed in the
literature situation of a uniform colormagnetic field.
I emphasize that in the present model  this field is uniform in the
intrinsic, "body fixed" frame.   For simplicity I will
first work in 2+1 dimensions  and will  try to extend to 3+1 in the
next section. I set
\beq
 a^i_\alpha (x) =
 \frac{1}{2}b_{\alpha} \epsilon_{ij}x^j
\label{cb1} \eeq
where the space indices $i,j$ presently run over the values 1 and 2.
In two space dimensions one can take $b$ diagonal in  color
since the transformation diagonalizing it is a part of $U$'s in
 (\ref{rgd}).
Explicit expression for $J$ is easily obtained
in this case from the known  Green's function
of a \Sch  equation in a constant magnetic field, cf. Ref.\cite{fey},
\beq
J_{\alpha \beta}(x,y) = \frac{1}{4\pi} e^{i(gb_{\alpha \beta}/2)
\epsilon_{ij} x^i y^j} \int_0^{\infty} \frac{ds}{\sinh s}
e^{-(|g b_{\alpha \beta}|/4)(x-y)^2 \coth s} \eeq
where $b_{\alpha \beta} = b_\alpha - b_\beta$. For $x = y$ this expression
is independent of color indices. It must be regulated to prevent the
divergence, e.g.
\beq
 \frac{g^2 N}{2\pi} \int_{s_0}^{\infty} \frac{ds}{\sinh s},\eeq
where $s_0$ is a regularization
cutoff. Although $J(x,y)$ does not depend only on
the distance $|x-y|$ the \Sch equation
\rf{peq}) with $P$ and $Q$ based on such $J$
is translationally invariant. Shifting  the coordinates by say a vector
$h$ and simultaneously performing a gauge transformation of the wave function
$c_{\alpha \alpha} \rightarrow c_{\alpha \alpha}\exp\left[i(gb_{\alpha}/2)
\epsilon_{ij}h^i (x_0^j - y_0^j)\right]$ leaves  Eq.\rf{peq})
invariant.  The integral in  the expression for $J(x,y)$
can be expressed in terms of the Bessel function
$K_0(|g(b_{\alpha} - b_{\beta}|z^2/4)$  with $z = x - y$ and for
 $|g (b_{\alpha} - b_{\beta})| z^2 \rightarrow \infty$
it has the following asymptotic form
\beq
\frac{1}{4\sqrt{\pi}}(2|g(b_\alpha - b_\beta)|)^{-1/4} \exp\left[ -
 |g(b_\alpha - b_\beta)| z^2/4\right]. \eeq
For finite values of  $g|b_{\alpha} - b_{\beta}|$
it decreases as a Gaussian at large separations
$z$. This should lead to a similar decrease of
the eigenvalues of \rf{peq})  -- an entirely unsatisfactory behavior
as far as the confinement is concerned.
In the next Section it will be seen that the situation may be
different in the large N limit.

\section{ Large N Random Colormagnetic Fields.}

As mentioned in the Introduction
rigid gauge field rotations should be relevant
for QCD in the large N limit  where it is expected that a
master field configuration dominates the vacuum, \cite{wit}.
As in the case of a condensate
such a configuration can not be just some fixed
gauge field potential $A^i_a(x)$.
It must be allowed to undergo free gauge "rotations"
exactly as $a^i$ in
Eq.(\ref{rgd}) since the gauge invariance is not expected to be  broken
in the large N limit. The dynamics of these rotations can not be "frozen"
and must be described by the gauge rotor Hamiltonian considered in
 Section 2. These "rotations" induce an interaction
between quarks as was shown in Section 3 for static quarks and will be
demonstrated for dynamical quarks in Section 6 below where it will
 also be shown that in addition
$a^i$ appears as a background field in the Dirac operator.

Another  important consideration is that for large N there is a large
number of degrees of freedom operating at each space point which
introduces statistical elements in the theory, cf. Refs.
\cite{mat,ran,bor}.
Experience with this limit for simple systems indicates that
two types of gauge invariant physical operators  should exist, analogoes to
macroscopic and microscopic observables in thermodynamics. The former
depend on finite (relative to N) number of dynamical variables
and involve sums over all labels of the degrees of freedom,
i.e. the color indices. A simple example is
$a^i_{\alpha\beta}a^i_{\beta\alpha}$, etc. Operators without such
summations, e.g., $a^i_{\alpha\beta}$ with fixed $\alpha$ and $\beta$
belong to the second  type which must be regarded as
microscopic observables
 like ,e.g., a coordinate of a particle or a single
spin variable in thermodynamic systems. The fluctuations of the macroscopic
operators are suppressed and expectations of their products factorize at
$N = \infty$. This is not so for microscopic observables.

On the basis of these considerations one can adopt the following point of view.
After allowing for free gauge rotations according to (\ref{rgd}), i.e.
after transformation to the  body-fixed frame, one should
 consider $a^i(x)$ as  static
 random matrix functions described by a probability
distribution $P[a^i(x)]$. This distribution can be determined following
the ideas of the random matrix theory, cf.
 Ref.\cite{ent}. To this end one should introduce the amount  of information
 (negative entropy)
\beq
I\left\{P\left[ a \right]\right\} =
 \int D\mu[a^i(x)] P[a^i(x)] ln P[a^i(x)]  \label{inf}
\eeq
associated with the $P[a^i(x)]$.
Minimizing $I\{P[a]\}$ subject to suitably chosen constraints on
macroscopic-like variables should determine the least biased distribution
$P[a^i(x)]$. As in statistical mechanics
the large N factorization should   then simply appear as a consequence of
the central limit theorem.

There are two crucial questions which need to be answered
in following this procedure - what is the appropriate measure in
the integral (\ref{inf}) and  what are the variables which should
be constrained. I hope to address the general answer to
these questions in the future work. Presently I will illustrate
how the procedure can be put to work for a uniform colormagnetic field,
Eq.(\ref{cb1}).

   In the limit of  large N
only the first terms in the expressions for $P$ and $Q$ above should
be retained and the \Sch  equation \rf{peq}) for diagonal components
of the string amplitude becomes
\beq
-2g^2\sum_\beta J_{\alpha \beta} (x_0,y_0) c_{\beta \beta} =
(E - E^0_\alpha) c_{\alpha \alpha}, \label{meq} \eeq
where $E^0_\alpha = 2g^2 \sum_\mu J_{\alpha \mu}(x_0,x_0)$.
The non diagonal string amplitudes decouple and satisfy a trivial equation
$(E^0_\mu + E^0_\nu)c_{\mu \nu} = 2 E c_{\mu \nu}$ the eigenvalues of which are
simply the sums of the selfenergies. Without careful treatment of long and
short distance regularization in the large N limit one can not reliably
discuss these eigenvalues taken separately and I will concentrate
on Eq.\rf{meq}).  Using translational invariance and
writing this equation for $x_0 = 0$ and $y_0 = z$ one obtains
\beq
\sum_{\beta=1}^N \int_0^{\infty} \frac{ds}{4\pi \sinh s}
e^{-(|g (b_{\alpha} - b_{\beta})|/4) z^2 \coth s} c_{\beta \beta} =
- \frac{E}{2g^2} c_{\alpha \alpha}.\label{teq}  \eeq
as the large N limit of the \Sch equation for a static quark --
antiquark pair in the rigid  gauge
configuration corresponding to a uniform colormagnetic field in 2+1
dimensions. One must still specify the large N scaling of various quantities
which enter this equation. Provided each term in the sum on the left hand
side is of the same order of magnitude  I get the standard
scaling of the coupling constant  requiring that $g^2 N$ is held fixed.
In the exponential of the integrand one can then extract the finite combination
$\bar g = g\sqrt{N}$. The problem is then to determine the
scaling and in general the entire distribution of
the field components  $b_{\alpha}/\sqrt{N}$. Regarding the behavior at
large separations $z$ one notes that if
the limit of $N \rightarrow \infty$ is taken first in such a way that
the differences $|b_{\alpha} - b_{\beta}|/\sqrt{N}$ decrease then the Gaussian
decay can possibly be prevented.

I use this example to  demonstrate how the ideas about the
statistical nature of the large N limit can be used to determine the
distribution of the components $b_\alpha/\sqrt{N}$.
I consider what happens with the Wilson loop
$W(C) = \frac{1}{N}<Tr P \exp(ig\oint_C A^idx_i)>$ in the present theory.
Choosing the loop perpendicular to the time axis, inserting \rf{rgd})
in $W(C)$, using its  gauge invariance and the explicit form \rf{cb1}) of
$a^i$  one finds
\beq
 W(C) = \frac{1}{N}\sum_\alpha e^{ig b_\alpha S} = \int_{-\infty}^\infty
 db \rho (b) e^{igbS}
\eeq
where $S$ is the area of the loop and $\rho (b) =
(1/N)\sum_{\alpha=1}^N \delta (b - b_\alpha)$ is the density of the field
components. In the large N limit $\rho(b)$ can be approximated
by a smooth function provided the range of variations of b does
not grow with N. Assuming this one easily finds simple expressions for
$\rho (b)$ , e.g. Lorenzian
\beq
\rho (b) = \frac{b_0 \sqrt{N}}{\pi (b^2 + Nb_0^2)} \label{lor} \eeq
which lead to the area law dependence of the Wilson loop,
$W(C) = \exp (-\bar{g}b_0 S)$. The combination $\bar{g} b_0$ plays the role of
the string constant. The placing of $N$'s in \rf{lor}) was chosen in such
a way as to have this constant finite for $N \rightarrow \infty$.

The choice \rf{lor}) is the simplest possible.
Any meromorfic function $\rho(b)$ with poles in the upper plane
will give the area law with the string tension controlled by the
position of the pole closest to the real axis. One can also take functions
with other type of singularities in the upper complex plane, etc. The simple
choice \rf{lor}) gives area law for any S,
missing entirely the asymptotic freedom behavior
at small S. One can attempt to correct this by choosing more
involved expressions for $\rho$. A  much more serious problem
is that the space oriented Wilson loop may not be a good measure
of confining properties in the model where one has
never worried about the  Lorenz invariance.

Adopting any  form of the "single component"  density $\rho(b)$
still leaves the distribution of the values of $b_\alpha$ needed
in, e.g., Eq. \rf{teq}) largely undetermined. Using
 statistical concepts described in the beginning of this
Section one should  view $b_\alpha$'s as random quantities
and introduce their joint probability distribution
$P(b_1, b_2,...,b_N)$ which should be such
that $\rho(b)$ is reproduced but is
otherwise  maximally random, i.e. contains least amount of  information.
The question immediately arises as to whether $\rho(b)$ is the only
quantity which should constrain $P(b_1,...,b_N)$ and what is the complete
set of such constraints. In the absence of  general answers
I take  $\rho(b)$ controlling $W(C)$ as an example
and determine the distribution $P(b_1,...,b_N)$ by
minimizing the appropriate negative entropy (information)
with this constraint.

The quantities $b_\alpha$'s are eigenvalues of a
hermitian, in general complex matrix. The information
content of a probability distribution $P(b_1,...,b_N)$ of such eigenvalues
is a well studied question, cf. Ref.\cite{ent}. It is
\beq \int d\mu[b] P(b_1,...,b_N) ln P(b_1,...,b_N)\label{ent}\eeq
where the measure is $d\mu[b] = const \prod_{\alpha > \beta} |b_\alpha
- b_\beta|^2
db_1 db_2...db_N$,  reflecting the repulsion of the eigenvalues. Minimizing
\rf{ent}) under the condition of a given $\rho (b) =
(1/N)\sum_\alpha (b - b_\alpha)$ one finds
\beq
P(b_1,...,b_N) = const \;\exp \left(\sum_{\alpha\neq\beta}^N ln|b_\alpha -
b_\beta| - 2N\sum_\alpha^N \int_{-\infty}^\infty ln|b_\alpha - b^\prime|
\rho (b^\prime ) db^\prime\right).\label{dis} \eeq
 Using, e.g. Eq. \rf{lor}) for $\rho (b)$ this expression becomes explicitly
\beq
P(b_1,...,b_N) = const\; \exp \left(\sum_{\alpha\neq\beta}^N ln|b_\alpha -
b_\beta| - N\sum_\alpha^N ln (b_\alpha^2 +Nb_0^2) \right).\label{pbb} \eeq
The constant in front of this expression must assure the normalization of
$P$ and can be calculated by the methods described in Ref.\cite{meh}.
Using the standard  interpretation  of $P(b_1,...,b_N)$  as a partition
function of a fictitious Coulomb gas
 one can say that the "particles" $b_\alpha$
are "repelled" from each other by the first
term in its exponential but are kept within the interval  $b_0$ by the second
term representing the interaction with the background "charge" distributed
according to $\rho(b)$. The average distance $|b_\alpha - b_\beta| \sim b_0/N$
becomes very small in the large N limit. The \Sch  equation \rf{teq}) is now
a random matrix equation with the probability distribution of its elements
controlled by the $P(b_1,...,b_N)$ above. The actual numerical
solution of this equation is now in progress.

In a similar way one can consider rigid gauge configuration representing
uniform colormagnetic field in $3 + 1$ dimensions. This field in the intrinsic
frame corresponds to the choice
\beq
a^i(x) = \frac{1}{2} \epsilon_{ijk} b^j x^k \label{cb2} \eeq
where  now however  the three color matrices $b^i$ in general cannot be
assumed diagonal. For such non-diagonal colormagnetic field the inversion
of the moment of inertia operator  $-(1/4g^2)[d^i,[d^i,\omega]]$
requires a solution of a matrix differential equation. This equation
simplifies considerably if $b^i$'s are nonetheless restricted to be diagonal,
$b^i_{\alpha \beta} = \delta_{\alpha \beta} b^i_\alpha$. Then
the equation \rf{dmi}) is still valid with the index $i$ now running from
1 to 3. In the following equation \rf{grf}) for the Green's function  one
should just replace $(b_\alpha - b_\beta)\epsilon_{ij}x^j$ by
$\epsilon_{ijk}(b^j_\alpha -
 b^j_\beta)x^k$. The expression for this Green's function is known
 and one can repeat all the
steps leading to the  static quark--antiquark
\Sch equation which is the analog of Eq.\rf{teq}) in 3+1 dimensions.

Turning again to the Wilson loop one finds in this case
 $\oint_C a^i dx_i =  b^j S_j$ with $S_j = (1/2) \epsilon_{jki}
\oint_C x^k dx_i$ so that $\left( \oint_c a^i dx_i \right)_\alpha =
b_\alpha S \cos \theta _\alpha$ (no sum over $\alpha$) where
$S = \sqrt{\sum_i (S^i)^2},  b_\alpha = \sqrt {\sum_i (b^i_\alpha)^2}$
and $\theta_\alpha$ -- the angle between the vectors $b^i_\alpha$ and $S^i$
at a given $\alpha$. S is the area of the loop when it is planar and
is related to the minimal area in general.
The Wilson loop is
\eqna
W(C) & = & \frac{1}{N}<\sum_\alpha e^{igSb_\alpha \cos \theta_\alpha}>
 = \nonumber \\
   & = & \frac{1}{N}\sum_\alpha 2\pi \int _{-1}^{1} d(\cos \theta_{\alpha})
e^{igSb_\alpha \cos \theta_\alpha} = \label{wl3} \\
    & = & \frac{4\pi}{gS} \int_0^{\infty} \frac{db}{b} \rho(b) \sin (gbS),
 \nonumber \eqne
where $\rho (b) = (1/N) \sum_\alpha \delta (b - b_\alpha)$ is the density
of the positive lengths of the color components of the vector
$b^i$. In \rf{wl3}) I have performed the angle averaging which must be
present in the vacuum wavefunction.
One can easily choose $\rho (b)$, e.g. the square of Lorentzian
\beq
\rho(b) = \frac{4 b_0 b^2\sqrt{N}}{\pi (b^2 + Nb_0^2)^2} \label{rh2} \eeq
which gives  the area law $W(C) = 4\pi\exp (-\bar{g}b_0S)$. This choice is
again not unique  and gives the area law for any $S$. It has a powerlike
tale as opposed to the perturbative Gaussian.

The statistical arguments for finding the entire distribution
of $b^i_\alpha$  can be used in $3+1$ dimensional
case as well with the difference that in this case the density of the lengths
of the vectors $b^i_\alpha$ is fixed by, e.g. Eq.\rf{rh2}) and their directions
are distributed isotropically.

\section{Dynamic Quarks}

Dynamic quarks can be easily included in the rigid gauge rotation model.
For this I define quark fields in the "rotating frame", $q = U\hat{q}$,
use Eq. (\ref{rgd}) in the second term of the QCD \H (\ref{ham})
and replace the first term in it by (\ref{hrot}). Using moreover
 the Gauss law constraint (\ref{gl}) I can write the original
QCD Hamiltonian (\ref{ham}) in the rigid gauge rotor limit as
expressed in terms of the quark fields only,
\beq
H_{rot} = \frac{1}{2}\int d^3 x d^3 y
\hat{\rho}_a (x) I_{ab}^{-1}(x,y,[a^i]) \hat{\rho}_b (y)
+\int d^3 x \hat{q}^{+}(x)[\alpha^{i}(p^{i}-g
a^{i}(x))+\beta m]\hat{q}(x), \label{hrt1} \eeq
where $\hat{\rho}_a = \hat{q}^{+}\lambda ^a\hat{q}$
are the color quark densities in the rotating frame.
The \H $H_{rot}$  describes quarks with gauge strings attached to them,
i.e. $q(x)$  are multiplied by $U^{+}(x)$. They  move in an external
colormagnetic field described by the vector potential $a^i(x)$ and
 interact
via an instantaneous  interaction  $I^{-1}_{ab}(x,y,[a^i])$ also
depending on $a^i$ via Eq.(\ref{mom}). The simultaneous appearance of
the rigid gauge field configuration both as a background and as "inducing"
the quark-quark interaction is ultimately a consequence of the
gauge invariance which requires that non dynamical rigid gauge fields
appear only in the form (\ref{rgd}).

$H_{rot}$ is gauge invariant since the operators $\hat{q}$ are. Moreover
this Hamiltonian should only be used in the color singlet sector of the
theory since I have used the Gauss law to derive it.

For the vanishing $a^i$ the \H $H_{rot}$ describes free quarks interacting
Coulombically. Also for a general non zero $a^i$
$H_{rot}$ should be regarded as the QCD analogue
of the QED \H in which only the instantaneous Coulomb
interaction between the charges has been retained.
Indeed the analogue of the rigid gauge "rotations"  (\ref{rgd}) in QED
is $A^i(x,t) = a^i (x) + \partial ^i \chi(x,t)$
with abelian $U = \exp (ig\chi(x,t))$, fixed rigid $a^i (x)$ and dynamical
$\chi (x,t)$. Repeating the steps leading to \rf{hrt1}) one will
derive in the QED case the Coulomb interaction between the charge densities.

Regarding the possible role of $a^i(x)$ as the master field in the large N
limit
one has in $H_{rot}$ a way in which this field should enter the quark
sector of the theory, i.e. serving both as a background field and
perhaps somewhat surprisingly also  controlling
the quark interaction. Following the developments of Section 5 this
field should be regarded as random. The appearance of a random interaction
between the quarks means that a possible mechanism for confinement
of dynamical quarks in the large N limit could be
related to the localization of their relative distances.
The possible connection between
 confinement and localization has already been mentioned
in the past but usually in the context of a random background field and not
with random interactions as appear in the present model.

The \H \rf{hrt1}) takes exact account of the gauge symmetry.
One must however also worry about global symmetries.
For any N the \H $H_{rot}$ may serve as a possible basis for various
phenomenological
developments. Both for this matter and conceptually  one must face the
issue that allowing for an arbitrary x-dependence of various color components
of $a^i$  in \rf{hrt1}) leads to breaking
of important symmetries such as translational, rotational, Lorenz,
time reversal, and various discrete symmetries. Of course the breaking
of continuous space symmetries is not uncommon in phenomenology,
e.g. the bag model, the quark potential model, the Skyrme model, etc.
 Symmetries can be restored by
considering all configurations translated by the symmetry and integrating
over them using collective coordinates. This of course
applies to both continuous and discrete symmetries.
In the absence of the guidance from the
symmetries a more dynamical criterion for fixing
$a^i$ seems to be the condition of lowest  energy.
This leads naturally to a  generalization of the variational approach
of Section 4 in which the  variational energy
should be replaced by the ground state energy  $E_0[a^i(x)]$
of (the suitably regularized)  $H_{rot}$  found for a given $a^i (x)$.
Should the solution $a^i$
 break a global symmetry, the symmetry "images" of this $a^i$
will also be solutions and one should "sum" over all of them in a standard
way thereby restoring the symmetry.
This variational approach may be combined with the Hartree-Fock
method which should allow to calculate $E_0 [a^i(x)]$ approximately. The
Hartree-Fock approximation for fermions was shown to be consistent with
the large N approximation, c.f., Ref.\cite{sal}.
In a combined approach one should form for fixed $a^i$
an expectation value of
$H_{rot}$ with respect to a trial state of a chosen
color singlet  configuration of quarks
(e.g., vacuum, baryon, etc) which must be a product state, i.e. such that
the expectation values with respect to this state have a non interacting
factorized form,i.e.,
\eqna
 <\hat{\rho}_a (x) \hat{\rho}_b (y)> &=& \lambda^a_{\alpha \beta}
\lambda^b_{\gamma \delta}(
<\qh_\alpha^{+}(x)\qh_\beta(x)><\qh_\gamma^{+}(y)\qh_\delta(y)> -\nonumber
\\
&-&<\qh_\alpha^{+}(x)\qh_\delta(y)><\qh_\gamma^{+}(y)\qh_\beta(x)>).
\label{hf}  \eqne
For  a global color singlet state
$<\qh_\alpha^{+}(x)\qh_\beta(y)> = \delta_{\alpha \beta}\rho (y,x)$, with
$\rho(x,y) = \frac {1}{N}\sum_{\gamma}<\qh_\gamma^{+}(y)\qh_\gamma(x)>$
and therefore
\beq
<\hat{\rho}_a (x) \hat{\rho}_b (y)> = -2\delta_{ab}\rho(x,y)\rho(y,x)
\label{fck} \eeq

Following the standard Hartree-Fock routine, cf., Ref.\cite{hfr},
the single quark density matrix can be expanded
in terms of a complete set of functions, i.e.,  $\rho (x,y) =
\sum_n f_n \psi_n (x) \psi_n^* (y)$. At this stage it is customary
to add the so called Slater determinant condition
which means that the single quark states $\psi_n$ have sharp occupations
 $f_n = 0$ or $1$.  This condition and its
compatibility with the large N limit and Lorenz invariance were discussed in
Ref.\cite{bor}. Adopting this condition, using \rf{fck}) in forming the
 expectation
 $E_{HF}[\rho(x,y),a^i(x)] = <H_{rot}>$ and varying with the respect to
$\psi_n(x)$'s with constraints on their
normalization one obtains the Hartree-Fock equation
\beq
[\alpha^{i}p^{i}+\beta m]\psi_n(x)
- 2\int d^3 y I_{aa}^{-1}(x,y,[a^i]) \rho (x,y)\psi_n(y)
 = \epsilon_n \psi_n(x), \label{feq} \eeq
which appears here as a selfconsistent
Dirac equation for the quark wave functions $\psi_n(x)$.
Note that  in this equation $a^i$ disappeared from the
Dirac operator and enters only the interaction $I_{aa}^{-1}(x,y)$.
 Solutions of \rf{feq}) determine the optimum $\rho$ and thus
$E_{HF}$ for a given $a^i$.

 Hartree-Fock equation similar to
\rf{feq}) has  been investigated in the  1+1 dimensional QCD, \cite{bor}
 although
there for obvious reasons the field $a^i$ was absent and $I^{-1}(x,y)$ was
simply $|x - y|$.   Both
the t'Hooft meson spectrum and  the baryon soliton solutions
were found in this approach
in 1+1 dimensions. For small quark masses the baryon was the realization of the
skyrmion described in the quark language. If
successful the approach based on Eq.\rf{feq}) can possibly lead to similar
results in 2+1 and 3+1 dimensions.

\section*{Acknowledgement}
Useful discussions with S. Elizur, S. Finkelstein, J. Goldstone, K. Johnson,
A. Kerman,  F. Lenz, M. Milgrom, J. Negele, J. Polonyi, E. Rabinovici,
S. Solomon and H. Weidenm\"uller are acknowledged gratefully. Special
acknowledgement is to A. Schwimmer for patient teaching and
answering my questions and for critical reading of the manuscript. Parts of
this work were done during visits in Max-Planck-Institut fur Kernphysik
at Heidelberg and I wish to thank H. Weidenm\"uller for warm hospitality
 there.


\begin{thebibliography}{999}
\bibitem{ins}A.A. Belavin, A.M. Polyakov, A.Z. Schwartz and Yu.S. Tyupkin,
Phys. Lett. B59 (1975) 85; G. t'Hooft, Phys. Rev. Lett. 37 (1976) 8;
Phys. Rev. D14 (1976) 3432; A.M. Polyakov, Nucl. Phys. B120 (1977) 429;
C. Callan, R. Dashen and D. Gross, Phys. Lett. 63B (1976) 334; 66B (1977)
375; Phys. Rev. D17 (1978) 2717; D19 (1979) 1826; S. Coleman, The Uses of
Instantons, Erice Lectures, 1977; E.V. Shuryak, Nucl. Phys. B203 (1982)
93,116,140; B214 (1983) 237;  D. I. Diakonov and V. Yu. Petrov, Nucl.
 Phys. B245 (1984) 259, B272 (1986) 457, B306 (1988)  809, B323 (1989) 53.

\bibitem{lar} t'Hooft, Nucl. Phys. B72 (1974) 451; B75 (1974) 461.

\bibitem{fra} E. Brezin, C. Itzykson, G. Parisi and J.B. Zuber,
Commun. Math. Phys. 59 (1978) 35.

\bibitem{wit} E. Witten, Nucl. Phys. B160 (1979) 57;
S. Coleman, 1/N, Erice Lectures, 1979, in S. Coleman, Aspects of Symmetry,
Cambridge University Press, 1985.

\bibitem{mgm} Yu.M. Makeenko and A.A. Migdal, Phys. Lett. 88B (1979) 135;
89B (1980) 437(E); 97B (1980) 253; Nucl. Phys. B188 (1981) 268.

\bibitem{egk} T. Eguchi and H. Kawai, Phys. Rev. Lett. 48 (1982) 106;
G. Bhanot, U. Geller and H. Neuberger, Phys. Lett. 113B (1982) 47;
115B (1982) 237; V.A.Kazakov and A.A. Migdal Phys. Lett. 116B (1982) 423;
A.A. Migdal, Phys. Lett. 116B (1982) 425.

\bibitem{lat} K. Wilson, Phys. Rev. D14 (1974) 2455; in Methods in Field
Theory, ed. R. Balian and J. Zinn-Justin, 1981, p. 261;
M. Creutz, Phys. Rev. D21 (1980) 2308.

\bibitem{ind} V.A. Kazakov and A.A. Migdal, preprint LPTENS-92/15 (1992).

\bibitem{str} J. Kogut and L. Susskind, Phys. Rev. D11 (1975) 395;  J.
Kogut, Rev. Mod. Phys. 51 (1979) 659.

\bibitem{pol} A.M.Polyakov, Phys. Lett. 59B (1975) 82; Nucl. Phys. B120
(1977) 429; A.M. Polyakov, Gauge Fields and Strings, Harwood, New York,
1987.

\bibitem{top} G. t'Hooft, Nucl. Phys. B138 (1978) 1; B153 (1979) 141;
 B238 (1981) 406.

\bibitem{svz} M.A. Shifman, A.I. Vainstein and V.I. Zakharov,
Nucl. Phys. B147 (1979) 385, 448, 519.

\bibitem{cop} H.B. Nielsen, Phys. Lett. 80B (1978) 133;
H.B.Nielsen and P.Olesen, Nucl. Phys. B144 (1978) 376; B160 (1979) 380;
P. Olesen, Phys. Scripta, 23 (1981) 1000, Nucl. Phys. B200[FS4] (1982) 381.

\bibitem{bro} H.C. Pauli and S.J. Brodsky, Phys. Rev. D32 (1985) 1993;
D32 (1985) 2001; K. Hornbostel, S.J. Brodsky and H.C. Pauli, Phys. Rev.
D41 (1990) 3814; R.J. Perry, A. Harindranat and K. Wilson,
Phys. Rev. Lett. 65 (1990) 2959; S. Brodsky, H.C. Pauli, G. McCarter
 and S. Pinsky, "The Challenge of Light-Cone Quantization of Gauge Field
Theory", OHSTPY-HEP-T-92-005; F. Lenz, S. Levit, M. Thies
and K. Yazaki, Ann. Phys. 208 (1991) 1.

\bibitem{jon} K.Johnson, L. Lellouch and J. Polonyi, Nucl. Phys. B367
(1991) 675.

\bibitem{adl} H.Pagels and E. Tomboulis, Nucl. Phys. B143 (1978) 485;
 S.L.Adler, Phys. Rev. D33 (1981) 2905, D24 (1981) 1063(E), Phys. Lett.
 110B (1982) 302, Nucl. Phys. B217 (1983) 381, S.L. Adler and T. Piran,
  Rev. Mod. Phys. 56 (1984) 1.

\bibitem{yaf} L.G. Yaffe, Rev. Mod. Phys. 541 (1982) 407, F.R. Brown and
 L.G. Yaffe, Nucl. Phys. B271 (1986) 267, T.A. Dickens et al, Nucl.
 Phys. B309 (1988) 1.

\bibitem{dua} M. Baker, James S. Ball and F. Zachariasen,  Phys.
Rev. D37 (1988) 1036; D38 (1988) 1926; D41 (1990) 2612.

\bibitem{mat} C. Itzykson and J.B. Zuber, J. Math. Phys. 21 (1980) 411.

\bibitem{ran} O. Haan, Phys. Lett. B106 (1981) 207; Yu. M. Makeenko, lectures
in "Gauge Theories of the Eighties",  Arctic School Proc., Acaslompolo,
Finland, 1982, Eds. R. Ratio and J. Lindfors;  P. Olesen, Nucl. Phys. B200
(1982) 381.


\bibitem{bor} S.Levit, Strange Solutions of Mean Field Equations
for Large N QCD in 1+1 Dimensions, 28 International Winter
 Meeting of Nuclear
Physics, Bormio (Italy), 1990.

\bibitem{sal} L.Salcedo, S.Levit and J.W.Negele, Nucl. Phys. B361 (1991) 585.


\bibitem{lee} N.H. Crist and T.D. Lee, Phys. Rev. D22 (1980) 939; T.D. Lee,
Particle Physics and Introduction to Field Theory, Harwood, New York,
1981, Ch.18.

\bibitem{jac} R. Jackiw and J. Goldstone, Phys. Lett. 74B (1978) 81.

\bibitem{sim} Yu.Simonov, Sov. J. Nucl. Phys. 41 (1985) 835,1014,
Nucl. Phys. B307 (1988) 512.
\bibitem{dos} H.D. Dosch, Phys. Lett. 198B (1987) 177; H.D. Dosch and Yu.
Simonov, Phys. Lett. 205B (1988) 339.

\bibitem{man} S. Mandelstam, Phys. Lett. 53B (1975) 476, Phys. Rep.
23C (1976) 245; t'Hooft, CERN report and papers quoted above,
 J. Kogut and L. Susskind, Phys. Rev. D11 (1975) 395.

\bibitem{sav} I.A. Batalin, S.G. Matinyan and G.K. Savvidi, Sov. J. Nucl.
Phys. 26 (1977) 214; Matinyan and G.K. Savvidi, B156 (1979) 1;
G.K.Savvidy, Phys.Lett. 71B (1977) 133; 130B (1983) 303;
Nucl. Phys. B246 (1984) 302;  H. Pagels and E. Tomboulis, Nucl. Phys.
B143 (1978) 485; B.Simon, Ann. Phys. 146, 209 (1983)

\bibitem{bjo} J.D.Bjorken, Elements of QCD; in: Lectures on
 Lepton--Nucleon Scattering and QCD; Prog. Nucl. Phys. Vol.4; eds.:
W.B.Arwood, J.D.Bjorken, S.J.Brodsky and R.Straynowski.



\bibitem{fey} R.P. Feynman and A.R. Hibbs,  Quantum Mechanics and Path
Integrals,   Ch.3-6, McGrow--Hill,1965.

\bibitem{ent} R. Balian, Nuovo Cimento, B57 (1968) 183, and in
 From  Microphysics To Macrophysics , vol. 1 , Springer -- Verlag,
 Berlin,  Heidelberg,  1991.

\bibitem{meh} M.L.Mehta, Random Matrices, 2nd edition, Academic Press,1991.

\bibitem{hfr} P. Ring and P. Schuck, The Nuclear Many--Body Problem,
Springer-- Verlag, New York, Heidelberg, Berlin, 1980.
\end{thebibliography}
\end{document}